\documentclass[pre,twocolumn,amsmath,amssymb,nofootinbib,floatfix,superscriptaddress]{revtex4}

\usepackage{graphicx,bm,xcolor, bbold}
\usepackage{enumerate}
\usepackage{graphicx,bm}
\usepackage[normalem]{ulem} 

\makeatletter
\def\graphicscale{\twocolumn@sw{0.4}{0.4}}
\def\graphicthreescale{\twocolumn@sw{0.3}{0.4}}

\begin{document}

\title{Critical dynamics of three-dimensional
  ${\mathbb Z}_N$ gauge models \\ and the inverted XY universality class}

\author{Claudio Bonati} 
\affiliation{Dipartimento di Fisica dell'Universit\`a di Pisa and INFN,
        Largo Pontecorvo 3, I-56127 Pisa, Italy}

\author{Haralambos Panagopoulos} 
\affiliation{Department of Physics, University of Cyprus,
P.O. Box 20537, 1678 Nicosia, Cyprus}

\author{Ettore Vicari} 
\affiliation{Dipartimento di Fisica dell'Universit\`a di Pisa,
        Largo Pontecorvo 3, I-56127 Pisa, Italy}

\date{\today}

\begin{abstract}
We investigate the critical relaxational dynamics of the
three-dimensional (3D) lattice ${\mathbb Z}_N$ gauge models with $N=6$
and $N=8$, whose equilibrium critical behavior at their topological
transitions belongs to the inverted XY (IXY) universality class (this
is also the universality class of the continuous transitions of the 3D
lattice U(1) gauge Higgs models with a one-component complex scalar
field), which is connected to the standard XY universality class by a
nonlocal duality relation of the partition functions. Specifically, we
consider the purely relaxational dynamics realized by a locally
reversible Metropolis dynamics, as commonly used in Monte Carlo
simulations.  To determine the corresponding dynamic exponent $z$, we
focus on the out-of-equilibrium critical relaxational flows arising
from instantaneous quenches to the critical point, which are analyzed
within an out-of-equilibrium finite-size scaling framework. We obtain
the estimate $z=2.59(3)$. A numerical analysis of the equilibrium
critical dynamics give consistent, but less accurate, results. This
dynamic exponent is expected to characterize the critical slowing down
of the purely relaxational dynamics of all topological transitions
that belong to the 3D IXY universality class.  We note that this
result implies that the critical relaxational dynamics of the 3D IXY
universality class is slower than that of the standard 3D XY
universality class, whose relaxational dynamic exponent $z\approx
2.02$ is significantly smaller, although they share the same
length-scale critical exponent $\nu\approx 0.6717$.

\end{abstract}

\maketitle


\section{Introduction}
\label{intro}

Several phenomena in various physical contexts, ranging from
fundamental interactions~\cite{Weinberg-book,ZJ-book} to
condensed-matter physics~\cite{Anderson-book,Wen-book,Fradkin-book,
  Sachdev-book2,Sachdev-19,BPV-24-rev}, are described by models
with gauge symmetries.  In particular, many emergent collective phenomena
can be modelled by effective lattice gauge theories where fermionic or
bosonic matter fields are coupled to gauge variables belonging to
discrete or continuous gauge groups, such as ${\mathbb Z}_N$ and U(1)
or SU($N$) groups, respectively.  The properties of the various
thermodynamic phases and the nature of the phase transitions crucially
depend on the interplay between global and gauge symmetries, and, in
particular, on the role played by the gauge modes.

Some distinct broad classes of continuous transitions and critical
behaviors have been identified in lattice gauge models, see, e.g.,
Ref.~\cite{BPV-24-rev}.  Among them, a class of peculiar critical
phenomena comprises topological gauge transitions, see, e.g.,
Refs.~\cite{Sachdev-19,BPV-24-rev,Wegner-71,Kogut-79,FS-79,
  Savit-80,DH-81,NRR-03,BKV-02,BLS-20,BPV-24-decQ2,BPV-24-ncAH}. They
depart from the conventional critical behaviors described by the
Landau-Ginzburg-Wilson approach based on $\Phi^4$ field theories with
an order-parameter scalar field, see, e.g.,
Refs.~\cite{ZJ-book,Wilson-83,Fisher-75,PV-02}. Indeed such
transitions do not have any local gauge-invariant order parameter.
However, like conventional continuous transitions driven by a local
order-parameter field, the critical modes of purely topological
transitions develop a diverging length scale $\xi$, behaving as $\xi
\sim |T/T_c-1|^{-\nu}$ where $T$ is the temperature, $T_c$ is the
critical temperature, and $\nu$ is a universal critical exponent.
These critical modes are related to extended gauge-invariant objects,
for instance Wilson or Polyakov loops, due to the absence of a local
gauge-invariant order parameter.  Topological transitions in quantum
systems with gauge symmetries can be also driven by quantum
fluctuations in the zero-temperature
limit~\cite{Sachdev-19,RV-21,Sachdev-book2}.

Paradigmatic lattice gauge models undergoing topological transitions
are the three-dimensional (3D) lattice ${\mathbb Z}_N$ gauge models
and the 3D inverted XY (IXY) gauge model, defined later in
Sec.~\ref{model}. In particular, the topological transitions of the 3D
lattice ${\mathbb Z}_N$ models for sufficiently large $N$ ($N\ge 4$
for generic lattice systems with ${\mathbb Z}_N$ gauge
symmetry~\cite{BPV-24-rev}) share the same IXY universality class of
the 3D IXY gauge model. The 3D IXY model is related by duality to the
3D XY model with Villain action characterized by a global U(1)
symmetry~\cite{DH-81,NRR-03}, thus sharing the same length-scale
critical exponent $\nu\approx 0.6717$ of the 3D XY universality
class~\cite{PV-02,CHPV-06,Hasenbusch-19,CLLPSSV-20}. However, we
distinguish the XY and IXY universality classes, because the duality
mapping concerns only the free energy of the XY and IXY systems
without external fields, while the magnetic sector is completely
missed in the IXY gauge models.  It is also worth mentioning that
topological IXY transitions are also developed by lattice Abelian
Higgs models with both compact and noncompact U(1) gauge
variables~\cite{BPV-24-rev,BPV-24-decQ2,BPV-24-ncAH}. In particular,
the lattice U(1) Higgs models with a one-component complex scalar
field are effective models of superconductors, and their continuous
transitions are expected to belong to the IXY universality class, see,
e.g.,
Refs.~\cite{BPV-24-rev,DH-81,NRR-03,BPV-24-decQ2,BPV-24-ncAH,BPV-21-ncAH,
  HLM-74,HT-96,FH-96,BFLLW-96,OT-98,HS-00,KNS-02,MHS-02,Herbut-book,BPV-23b}.

Most studies of 3D lattice gauge models have investigated the
equilibrium static properties, focusing on the phase diagrams and
critical properties of the continuous transitions, see, e.g.,
Ref.~\cite{BPV-24-rev} and references therein. On the other hand, much
less is known about their critical dynamics, and, in particular, about
the critical slowing down associated with a local relaxational
dynamics, corresponding to the model A in the classification reported
in Refs.~\cite{HH-77,Ma-book,FM-06}.  Model A is the simplest
dynamical model governed by Langevin equations with a stochastic
force, which describes the dynamics of non-conserved fields relaxing
to an equilibrium condition determined by the partition function of
the model.  We recall that the critical dynamics in statistical models
is generally characterized by a power-law critical slowing down,
controlled by a universal dynamic exponent $z$~\cite{HH-77,Ma-book},
depending on the equilibrium universality class of the continuous
transition and the type of dynamics (whether it is purely relaxational
or there are conserved quantities). In the infinite-volume limit, the
time scale of the critical modes diverge as $\tau\sim\xi^z$
approaching the critical point, where $\xi$ is the diverging length
scale of the critical modes, while, in a finite volume and at the
critical point, they diverge as $\tau\sim L^z$, where $L$ is the
linear size of the system.

The purely relaxational dynamics of gauge field theories, driven by
corresponding Langevin equations, has been addressed in the context of
the stochastic quantization of gauge theories, see, e.g.,
Refs.~\cite{PW-81,Zwanziger-81,FI-82,NNOO-83,ZJ-86,ZZ-88}.  We also
mention some studies addressing particular aspects of the dynamics of
superconductors (see, e.g., Refs.~\cite{DFM-07, LVF-04, SBZ-02, AG-01,
  JKM-00, LWWGY-98, WJ-97, LMG-91}) and of hadronic matter close to
continuous phase transitions~\cite{RW-93}, involving models with
Abelian and non-Abelian gauge symmetries. In particular,
Refs.~\cite{DFM-07,LVF-04} analyzed the critical relaxational dynamics
associated with the $N$-component Abelian Higgs field theory, for
which a stable fixed point exists close to four dimensions only for a
sufficiently large number of components.  However, the study of the
critical dynamic phenomena in the presence of Abelian and non-Abelian
gauge symmetries is still far from being complete.

Only few results have been established for the critical dynamics of 3D
statistical models in the presence of a gauge symmetry.  In
particular, we mention those on the critical relaxational dynamics of
the 3D lattice ${\mathbb Z}_2$ gauge
model~\cite{BPV-25,BPV-25-z2equi,XCMCS-18,BKKLS-90}, whose equilibrium
critical behavior can be related to that of the standard 3D Ising
model by an exact nonlocal duality mapping of their free
energy~\cite{Wegner-71,Savit-80}, thus sharing the same length-scale
critical exponent $\nu\approx 0.630$ at their transition
points~\cite{PV-02}. However, a substantial difference has been found
in the power law describing the critical slowing down induced by local
relaxational dynamics. Indeed, the dynamic exponent
$z=2.610(15)$~\cite{BPV-25} associated with the local relaxational
dynamics of the 3D ${\mathbb Z}_2$ gauge model turns out to be
significantly larger than that of the local relaxational dynamics of
the standard Ising universality class, which is given by
$z=2.0245(15)$~\cite{Hasenbusch-20}.  Note that this difference of the
dynamic exponents does not contradict the exact equivalence of the
free energy of the 3D ${\mathbb Z}_2$ gauge and spin models, due to
the fact that the duality mapping is non local, therefore it cannot
relate their local relaxational dynamics, thus allowing them to
develop different dynamic power laws.  We believe that the extension
of these studies to other gauge-symmetric models may lead to a deeper
understanding of the various types of critical dynamics developed by
lattice gauge systems, in particular at topological transitions, such
as those belonging to the IXY universality class.

In the present work we address the critical dynamics of lattice gauge
systems undergoing IXY transitions, focusing on the purely
relaxational dynamics which, to our knowledge, has not yet been
studied. For this purpose we present a numerical study of the critical
dynamics of the lattice ${\mathbb Z}_N$ gauge models for $N=6$ and
$N=8$, whose equilibrium critical behaviors are known to belong to the
IXY universality class.  Specifically, we consider the purely
relaxational dynamics realized by a standard locally reversible
Metropolis dynamics, as commonly used in Monte Carlo (MC) simulations
of lattice models~\cite{Binder-76}. Our main goal is to estimate the
critical dynamic exponent $z$ of the local relaxational dynamics,
whose value is unknown. For this purpose, we focus on the
out-of-equilibrium relaxational flows of the gauge-invariant energy
density, arising from instantaneous soft quenches to the critical
point~\cite{PV-24}. Their analysis within an out-of-equilibrium
finite-size scaling (FSS) framework allows us to determine the dynamic
exponent $z$.  This approach was already exploited to study the
relaxational dynamics of the lattice ${\mathbb Z}_2$ gauge
model~\cite{BPV-25}, providing accurate results for $z$, improving
those obtained by the more standard analysis of the equilibrium
critical dynamics~\cite{BPV-25-z2equi,BKKLS-90}.

We anticipate our
final estimate $z=2.59(3)$ of the critical relaxational dynamic
exponent. This dynamic exponent is expected to characterize the
relaxational critical slowing down of all topological transitions that
belong to the IXY universality class, thus including those of the 3D
Abelian Higgs models (see, e.g., Ref.~\cite{LVF-04} for a discussion
of the relevance of the simplest relaxational dynamics for
superconducting systems effectively described by Abelian Higgs field
theories).  We also report a numerical investigation of the
equilibrium critical dynamics, obtaining consistent, although less
accurate, results.

We note that the above estimate of $z$ implies that the critical
relaxational dynamics of the 3D IXY universality class turns out to be
slower than that of the standard spin XY universality class, although
they share an analogous critical behavior in the thermal sector.
Indeed, the dynamic exponent $z$ associated with the purely
relaxational dynamics of the 3D XY universality class is significantly
smaller, given by $z=2.0246(10)$ (obtained in Ref.~\cite{AEHIKKZ-22}
by high-order perturbative computations, see also
Refs.~\cite{FM-06,AV-84,HH-77,HHM-72}).  In this respect, we stress
that the duality mapping between the 3D IXY and XY models is
nonlocal. Therefore, a local dynamics in the IXY model would
correspond to a nonlocal dynamics in the XY model, explaining why the
universal features of the local relaxational dynamics of the IXY
universality class differ from that of the standard XY model.

This paper is organized as follows.  In Sec.~\ref{model} we introduce
the 3D ${\mathbb Z}_N$ gauge models and the IXY model, summarizing
their known equilibrium critical properties. In Sec.~\ref{relprot} we
describe the purely relaxational Metropolis dynamics that we consider,
and the dynamic protocol adopted to simulate the critical relaxational
flows; we discuss the out-of-equilibrium behavior of the energy
density along the critical relaxational flow within an
out-of-equilibrium FSS framework, and present the numerical results
leading to the above-mentioned estimate of the relaxational dynamic
exponent $z$. In Sec.~\ref{equicritdyn} we also investigate the
equilibrium critical dynamics by analyzing the autocorrelation times
at the critical point, which further support the estimate of $z$
obtained from analysis of the out-of-equilibrium critical dynamics.
Finally, in Sec.~\ref{conclu} we summarize and draw our conclusions.

\section{The 3D lattice ${\mathbb Z}_N$ gauge model}
\label{model}

We consider 3D cubic-lattice ${\mathbb Z}_N$ gauge models defined by
the Hamiltonian~\cite{Savit-80,BPV-24-rev}
\begin{eqnarray}
  H  = - K \, \sum_{{\bm x},\mu>\nu} {\rm Re}\; \Pi_{{\bm x},\mu\nu},
  \label{zqgau}
  \end{eqnarray}
where $\Pi_{{\bm x},\mu\nu}$ is the plaquette operator 
\begin{eqnarray}
\Pi_{{\bm x},\mu\nu} = \lambda_{{\bm x},{\mu}} \,\lambda_{{\bm x}+\hat{\mu},{\nu}}
  \,\bar{\lambda}_{{\bm x}+\hat{\nu},{\mu}} \,\bar{\lambda}_{{\bm
      x},{\nu}}, 
  \label{plaquette}
\end{eqnarray}
constructed with the ${\mathbb Z}_N$-group link variables
\begin{eqnarray}  
\lambda_{{\bm x},\mu}= \exp\Bigl(i{2\pi n_{{\bm x},\mu}\over
  N}\Bigr),\qquad n_{{\bm x},\mu}= 1,\ldots,N,
\label{zqvar}
\end{eqnarray}
associated with the bond starting from site ${\bm x}$ in the positive
$\mu$ direction, $\mu=1,2,3$.  The Hamiltonian parameter $K$ plays the
role of inverse gauge coupling. In the following we set the
temperature $T=1$ without loss of generality, so that the partition
function reads $Z =\sum_{\{\lambda\}} e^{-H}$.  In our numerical
finite-size scaling (FSS) analyses we consider cubic-like systems of
size $L$ with periodic boundary conditions.

The 3D lattice ${\mathbb Z}_N$ gauge models undergo a phase transition
separating a high-$K$ deconfined phase from a low-$K$ confined phase.
This phase transition has a topological nature, without a local order
parameter.  However, one may identify a nonlocal order parameter,
related to the behavior of the so-called Wilson loops $W_C$, defined
as the product of the link variables along a closed contour $C$ within
a plane~\cite{Savit-80}.  Their size dependence for large contours
qualitatively changes at the transition point, passing from the
small-$K$ area law $W_C\sim \exp(- c_a A_C)$, where $A_C$ is the area
enclosed by the contour $C$ and $c_a>0$ is a constant, to the
large-$K$ perimeter law $W_C\sim \exp(- c_p P_C)$, where $P_C$ is the
perimeter of the contour $C$ and $c_p>0$ is a constant.

The transition point $K_c$ and critical exponents of the ${\mathbb
  Z}_N$-gauge topological transition can be inferred by using duality.
Indeed, they are dual to specific $N$-state clock
models~\cite{Savit-80,BCCGPS-14}, with $\mathbb{Z}_N$ spin variables $\exp(2\pi
i n_{\bm x}/N)$ (where $n_{\bm x}=1,...,N$) associated with the sites
of a cubic lattice, whose Hamiltonian is symmetric under global
${\mathbb Z}_N$ transformations.  The known
properties of the renormalization-group flow of the $N$-state clock
universality classes tell us that the critical behaviors of generic 3D
lattice ${\mathbb Z}_N$ gauge models with $N\ge 4$ must belong to the
3D XY universality class, or, more appropriately, to the 3D IXY
universality class~\cite{BPV-24-rev}.  We also note that the
lattice ${\mathbb Z}_4$ gauge model (\ref{zqgau}) represents a
particular case, for which the phase transition belongs to the
Ising universality class, because its partition function can be
written as that of two independent ${\mathbb Z}_2$ gauge
models~\cite{Korthals-78, BPV-24-decQ2} (this is analogous to what
happens in the standard ${\mathbb Z}_4$ clock model). However, this
result is specific to the gauge formulation with Hamiltonian
(\ref{zqgau}).  Generic lattice ${\mathbb Z}_4$ gauge models are
instead expected to undergo transitions belonging to the IXY
universality class~\cite{BPV-24-rev}.

The IXY universality class is generally represented by the topological
transition of the lattice IXY gauge model defined by the Hamiltonian
\begin{equation}
      H_{\rm IXY}= 
      \frac{\kappa}{2} \sum_{{\bm x},\mu>\nu} 
         (\nabla_\mu A_{{\bm x},\nu} - \nabla_\nu A_{{\bm x},\mu})^2 , 
\label{IXYham}
\end{equation}
where the sum is over all lattice plaquettes, $\nabla_\mu$ is the
forward lattice derivative, $\nabla_\mu f({\bm x}) = f({\bm
  x}+\hat{\mu}) - f({\bm x})$, and the field $A_{{\bm x},\mu}$ takes
only values that are multiples of $2 \pi$, i.e., $A_{{\bm x},\mu} = 2
\pi n_{{\bm x},\mu}$, with $n_{{\bm x},\mu} \in \mathbb Z$.  The
Hamiltonian $H_{\rm IXY}$ is invariant under the gauge transformations
$n_{{\bm x},\mu} \to n_{{\bm x},\mu} + m_{{\bm x}+\hat{\mu}} - m_{\bm
  x}$, with $m_{\bm x} \in \mathbb Z$.  The IXY gauge model undergoes
a continuous topological transition at $\kappa_c =
0.076051(2)$~\cite{NRR-03,BPV-21-ncAH}, when setting the temperature
$T=1$. The free energy of the IXY model is related by duality to that
of the XY model with Villain action~\cite{DH-81,NRR-03}.  This implies
that the divergence of the correlation length of the critical modes at
the topological transition is controlled by the same critical exponent
$\nu$ of the 3D XY universality class.

We remark that the identification of the universality classes of the
continuous transitions of the lattice ${\mathbb Z}_N$ and IXY gauge
models is done using the duality between the gauge and the spin
system, and that duality only maps the free energy and the related
thermal observables that can be obtained by its temperature
derivatives, such as the energy density. On the other hand, the
magnetic sector that is present in the spin XY universality class has
no counterpart in the gauge universality class.  Therefore, it is more
appropriate to distinguish the IXY universality class from the spin XY
universality class, see, e.g., Ref.~\cite{BPV-24-rev}.

In the following we focus on the 3D lattice ${\mathbb Z}_N$ gauge
models for $N=6$ and $N=8$. Their transition point was estimated in
Ref.~\cite{BCCGPS-14} by duality and MC simulations of the
corresponding ${\mathbb Z}_N$ clock models, obtaining $K_c =
3.00683(7)$ for $N=6$ and $K_c = 5.12829(13)$ for $N=8$.  As already
discussed, their critical behavior belongs to the IXY universality
class. In the following, we use the
estimates~\cite{CHPV-06,Hasenbusch-19,CLLPSSV-20}
\begin{equation}
  \nu=0.6717(1),\qquad \omega=0.789(4),
  \label{xyexp}
  \end{equation}
respectively for the length-scale critical exponent $\nu$ and the
exponent $\omega$ related to the leading scaling corrections.

\section{Critical relaxational flows}
\label{relprot}

To investigate the out-of-equilibrium critical dynamics at the
topological transition, we consider {\em soft} quench protocols around
the transition point~\cite{PRV-18,RV-21}, so that the
out-of-equilibrium evolution occurs within the critical region.  In
particular, we study the out-of-equilibrium critical relaxational flow
arising from an instantaneous quench of the gauge parameter $K$, from
the disordered phase, i.e., $K<K_c$, to the critical point $K_c$.

\subsection{Local relaxation by Metropolis upgradings}
\label{locrelmetro}

Our numerical study of the critical relaxational dynamics of lattice
${\mathbb Z}_N$ gauge models is based on equilibrium and
out-of-equilibrium MC simulations. We consider the local relaxational
dynamics realized by standard Metropolis upgradings of the link
variables~\cite{Binder-76}.  To upgrade a given spin variable
$\lambda_{{\bm x},{\mu}}$, we randomly take one of the two closest
group elements $\lambda_{{\bm x},{\mu}}'$. The transition probability
for a change of $\lambda_{{\bm x},{\mu}}$ is
\begin{equation}
  P(\lambda_{{\bm x},{\mu}}\to \lambda_{{\bm x},{\mu}}') = {\rm
    Min}[1,e^{-\Delta H}],
    \label{metroupg}
    \end{equation}
where $\Delta H$ is the variation of the lattice Hamiltonian
(\ref{zqgau}) when changing $\lambda_{{\bm x},{\mu}}\to \lambda_{{\bm
    x},{\mu}}'$. One MC time unit corresponds to a global sweep of a
single Metropolis upgrading proposal for each bond variable. We
consider a checkerboard upgrading scheme, where we separately upgrade
link variables along a given direction at odd and even sites.

In our Metropolis algorithm, the proposal is limited to the two
closest group elements of the actual link variables.  The acceptance
ratio of such Metropolis upgrading turns out to be relatively small at
the critical point $K_c$, about 2\%, for both ${\mathbb Z}_6$ and
${\mathbb Z}_8$ gauge models.  We also mention that this
nearest-values Metropolis proposal turns out to be more effective than
that of choosing among all group elements, which significantly reduces
the acceptance to about 0.6\% for the lattice ${\mathbb Z}_8$ model at
$K_c$, and also the effectiveness of the global upgrading.

We start all MC simulations from cold configurations, which represent
the optimal choice, given that the lattice ${\mathbb Z}_N$ gauge
models turn out to be quite ordered at their transition point
$K_c$. Indeed the average plaquette values $\langle \Pi_{{\bm
    x},\mu\nu} \rangle$ at $K_c$ turn out to be very close to one, for
example $\langle \Pi_{{\bm x},\mu\nu} \rangle\approx 0.974$ for $N=6$
and $\langle \Pi_{{\bm x},\mu\nu} \rangle\approx 0.985$ for $N=8$
[these values can be easily obtained from the critical energy densities
  reported in Eqs.~(\ref{ecn6}) and (\ref{ecn8})]. Then, we perform a
  sufficiently large number of sweeps to thermalize the system at
  $K<K_c$, and then study the critical relaxation flows starting from
  equilibrium configurations generated at $K$, as described below.

\subsection{Protocol for critical relaxational flows}
\label{dynprot}

The Gibbs ensemble of equilibrium configurations at $K<K_c$ are the
starting point for out-of-equilibrium critical relaxational flows at
the critical point $K_c$, which are realized by making the system
evolve using a purely local relaxational dynamics at
$K_c$~\cite{HH-77,Binder-76}, where the time $t$ is related to the
number of relaxational sweeps of the lattice variables, starting at
$t=0$.

In our study we monitor the out-of-equilibrium dynamics arising from
this quench protocol by the time dependence of the gauge-invariant
energy density
\begin{equation}
  E(t) = - {1\over L^3} \big\langle
  \sum_{{\bm x},\mu>\nu} {\rm Re}\; \Pi_{{\bm x},\mu\nu} 
      \big\rangle_t ,
   \label{defened}
   \end{equation}
defined as the average over the configurations obtained at a fixed
time $t$ along the relaxational flow.  This is the simplest
gauge-invariant quantity in lattice gauge models, and also the most
convenient quantity to monitor along the relaxational flow.  More
extended objects, like Polyakov or Wilson lines, turn out to be less
effective in this respect, essentially because they are subject to
much larger fluctuations, and therefore larger errors.

Since the dynamics is purely relaxational, the critical equilibrium of
finite-size systems is eventually recovered for large times, which
tend to be larger and larger with increasing $L$, due to the critical
slowing down. In particular, we must have that $E(t\to\infty)\to
E_c(L)$, where $E_c(L)$ is the equilibrium energy density at the
critical point. We are actually interested in the initial
out-of-equilibrium regime, which can be also described within an
out-of-equilibrium FSS framework controlled by the dynamic exponent
$z$ of the purely relaxational dynamics.

\subsection{Out-of-equilibrium scaling behavior}
\label{enesca}

We now summarize the main features of the expected out-of-equilibrium
behavior of the energy density along the critical relaxational flow
associated with the quench protocol outlined in
Sec.~\ref{dynprot}. For this purpose, we exploit an out-of-equilibrium
FSS framework~\cite{PV-24,BPV-25}.

According to the renormalization-group theory of critical phenomena,
the equilibrium energy density $E_e$,
\begin{eqnarray}
  E_e = - {1\over L^d} \langle \sum_{{\bm x},\mu>\nu} {\rm Re}\; \Pi_{{\bm
      x},\mu\nu}\rangle,
  \label{enedens}
\end{eqnarray}
around the transition point is expected to behave as~\cite{PV-02}
\begin{eqnarray}
&&E_e(K,L) \approx E_{\rm reg}(r) + L^{-(3-y_r)} {\cal E}_e(\Upsilon),
  \label{leadINGFSSene}\\
&&  r = K_c - K,\quad \Upsilon = r \,L^{y_r},\quad y_r=1/\nu.
  \label{Upsilondef}
\end{eqnarray}
The first contribution $E_{\rm reg}(r)$ represents an analytical
background; indeed it is a regular function of the deviation $r$ from
the critical point.  The second scaling term is the nonanalytic
contribution from the critical modes, and the scaling function ${\cal
  E}_e({\Upsilon})$ is expected to be universal, apart from a
multiplicative constant and a normalization of $\Upsilon$. The scaling
term is generally subleading with respect to the regular one, due to
the fact that $3-y_r>0$ at 3D continuous transitions.  Therefore, the
behavior of the energy density is dominated by the analytical
background. In a field-theoretical setting, we may interpret it as a
mixing with the identity operator.

In the case of the 3D XY universality class, and therefore also the
IXY universality class, the background analytical term provides the
leading contribution even if we consider the difference
\begin{eqnarray}
  E_{se}(r,L) = E_e(r,L) - E_{c,\infty}, \label{diffe}
\end{eqnarray}
where $E_{c,\infty}$ is the infinite-size energy density at the
critical point. By means of standard MC simulations at the critical
point for lattice sizes up to $L=44$ for $N=6$ and up to $L=64$ for
$N=8$, and large-$L$ extrapolations using the fit ansatz $E_c(L) =
E_{c,\infty} + c \,L^{-(3-y_r)}$, as predicted by
Eq.~(\ref{leadINGFSSene}), we obtained the estimates
\begin{eqnarray}
&E_{c,\infty}=-2.922610(7)\quad &{\rm for}\;\;N=6, \label{ecn6}\\
&E_{c,\infty}=-2.954746(4)\quad &{\rm for}\;\;N=8.\label{ecn8}
\end{eqnarray}
The scaling behavior of $E_{se}(r,L)$ can be straightforwardly
obtained from Eq.~(\ref{leadINGFSSene}). One can easily verify that,
due to the negative values of the specific-heat exponent $\alpha = 2 -
3\nu\approx -0.015$, the $O(r)$ contribution of the analytical
background of $E_{se}(r,L)$ is still dominant with respect to the
scaling term.  As put forward in Refs.~\cite{PV-24,BPV-25}, and
verified numerically, the problem of the analytical background gets
overcome along the out-of-equilibrium relaxational flow, when
computing the subtracted post-quench energy density
\begin{eqnarray}
  E_s(t,r,L) =  E(t,r,L)-E_{c,\infty}.
  \label{diffet}
\end{eqnarray}

The out-of-equilibrium FSS of the critical modes along the
relaxational flow is generally described by a further time scaling
variable $\Theta = t\,L^{-z}$.  As conjectured, and numerical
verified, in Refs.~\cite{PV-24,BPV-25}, the energy difference $E_s$
along the critical relaxational flow shows the asymptotic FSS behavior
\begin{eqnarray}
  \Omega(t,r,L) = L^{3-y_r} E_s \approx {\cal A}(\Theta,\Upsilon),
  \quad \Theta = t\,L^{-z}, \quad \label{erscal}
\end{eqnarray}
without the complications of analytical background contributions.  The
out-of-equilibrium scaling function ${\cal A}(\Theta,\Upsilon)$ is
expected to be universal, apart from a multiplicative constant and
trivial normalizations of the arguments.  The out-of-equilibrium
FSS of the energy density is expected to be approached with the same
$O(L^{-\omega})$ scaling corrections of the equilibrium FSS.

However, the $\Theta\to 0$ limit of the scaling function ${\cal
  A}(\Theta,\Upsilon)$ must be singular for any $\Upsilon>0$. Indeed,
the disappearance of the dominant $O(L^{-\alpha/\nu})$ equilibrium
contributions from the asymptotic behaviors at fixed $\Theta>0$
reemerges as a nonanalytical behavior of the scaling function ${\cal
  A}(\Theta,\Upsilon)$ in the $\Theta\to 0$ limit, i.e.,~\cite{PV-24}
\begin{equation}
  {\cal A}(\Theta,\Upsilon) \approx \Theta^{\alpha/\nu} f_1(\Upsilon) 
  + f_2(\Upsilon),
  \label{thetasmall}
\end{equation}
where $\alpha/\nu = 2 y_r - 3 \approx -0.022$.

In our numerical analyses we also consider the time $t(\Omega)$ as a
function of the rescaled subtracted energy density $\Omega$ (this is
well defined because $\Omega$ is a monotonic function of $t$ along the
critical relaxational flow).  Numerically, this can be easily
estimated by linearly interpolating the data of $\Omega(t,r,L)$ as a
function of $t$ (one may also use higher-order interpolations), using
a straightforward propagation of the statistical error.  One can
easily derive the corresponding out-of-equilibrium FSS from that of
the subtracted energy density, cf. Eq.~(\ref{erscal}), obtaining
\begin{eqnarray}
  t(\Omega,\Upsilon,L) = L^{z} \left[ F(\Omega,\Upsilon) +
    O(L^{-\omega})\right],
\label{generalansatzt}
\end{eqnarray}
where $\Upsilon= r \,L^{y_r}$.  As noted in Ref.~\cite{BPV-25}, it
turns out convenient to also consider the difference
\begin{eqnarray}
  \Delta(\Omega,\Upsilon_1,\Upsilon_2,L) &=& t(\Omega,\Upsilon_2,L) -
  t(\Omega,\Upsilon_1,L) \label{difft}\\ &\approx& L^z
  F_{\Delta}(\Omega,\Upsilon_2,\Upsilon_1), \nonumber
\end{eqnarray}
where $\Upsilon_i = r_i L^{y_r}$, because these differences are less
affected by scaling corrections.

\subsection{Numerical results}
\label{numres}

\begin{table}
\begin{tabular}{cccclc}
  \hline\hline
& fit  & $\Omega$ & 
  $L_{\rm min}$ & $\quad z$ & $\chi^2/{\rm
    dof}$ \\ \hline
$\Delta(\Omega,10,20,L)$ & a & 3.1  & 24  &  2.592(3)  & 0.7  \\
        &         &      & 32  &  2.598(5)  & 0.2  \\
        &         &      & 40  &  2.595(9)  & 0.2  \\
        &         & 2.85 & 24  &  2.595(3)  & 0.7  \\
        &         &      & 32  &  2.600(5)  & 0.6  \\
        &         &      & 40  &  2.602(8)  & 0.9  \\
        &         & 2.6  & 24  &  2.592(4)  & 1.1  \\
        &         &      & 32  &  2.590(7)  & 1.3  \\
        &         &      & 40  &  2.580(13) & 1.6  \\
  &  b    & 3.1  & 24  &  2.63(3)  & 0.4  \\
  &         & 2.6  & 24  &  2.63(3)  & 0.4  \\
  \hline

  $t(\Omega,10,L)$ & a &  2.8    &  32 & 2.534(2)  & 5.8  \\
                & &         &  40 & 2.551(5)  & 0.5  \\
                & &  2.4    &  32 & 2.553(2)  & 2.9  \\
                & &         &  40 & 2.561(4)  & 2.2  \\
                & &  2.1    &  32 & 2.567(5)  & 2.9  \\
                & &         &  40 & 2.561(7)  & 3.2  \\
  \hline

  $t(\Omega,20,L)$ & a &  3.8    &  40 & 2.550(4)  & 3.5  \\
                & &         &  48 & 2.563(8)  & 3.0  \\
                & &  3.1    &  40 & 2.570(4)  & 1.0  \\
                & &         &  48 & 2.574(8)  & 1.8  \\
                & &  2.6    &  32 & 2.586(4)  & 12  \\
  \hline

  $C_2$ & a & $_{3.4,2.6}$ & 40  &  2.556(2)  & 1.3  \\
  &       &    & 48  &  2.560(9)  & 0.4  \\
  &  & $_{3.1,2.4}$ & 40  &  2.573(4)  & 1.4  \\
  &       &    & 48  &  2.581(7)  & 1.6  \\
  \hline
\end{tabular}
\caption{Results for the dynamic exponent $z$ obtained by some fits of
  the data of $\Delta(\Omega,\Upsilon_1,\Upsilon_2,L)$ and
  $t(\Omega,\Upsilon,L)$ to the ansatz: (a) $a \, L^z$ and (b) $a
  \,L^z (1 + a_1 \,L^{-\omega})$ for the lattice ${\mathbb Z}_6$ gauge
  model, at fixed values of $\Omega$, using data for $\Upsilon=20$ and
  $\Upsilon=10$. We also report combined fits $C_2$ of the data of
  $t(\Omega,\Upsilon,L)$ for $\Upsilon=20$ and $\Upsilon=10$ (with
  $\Omega$ chosen so that they correspond to $\Theta\approx 0.02$). On
  the basis of these results, one may consider $z=2.595(35)$ as a
  final estimate for the lattice ${\mathbb Z}_6$ gauge model, where
  the central value is taken from the most stable fits of the
  differences $\Delta(\Omega,\Upsilon_1,\Upsilon_2,L)$, which are
  apparently subject to smaller scaling corrections, and the error
  allows for the spread of the results of the fits.}
  \label{fitsz6}
\end{table}

\begin{table}
\begin{tabular}{cccclc}
  \hline\hline & fit & $\Omega$ & $L_{\rm min}$
  & $\quad z$ & ${\chi^2/{\rm dof}}$ \\
  \hline $\Delta(\Omega,20,40,L)$
  & (a) & 2.2 & 24 & 2.575(3) & 2.2 \\
  & & & 32 & 2.585(5) & 0.7 \\
  & & & 40 & 2.585(7) & 1.0 \\
  & & 2.0 & 24 & 2.586(3) & 0.5 \\
  & & & 32 & 2.587(7) & 0.7 \\
  & & & 40 & 2.589(11) & 1.0 \\
  & & 1.8 & 24 & 2.593(3) & 0.8 \\
  & & & 32 & 2.598(6) & 0.8 \\
  & & & 40 & 2.606(10) & 0.5 \\
  & & 1.6 & 24 & 2.599(3) & 2.9 \\
  & & & 40 & 2.600(9) & 5.0 \\
  & (b) & 2.2 & 24 & 2.63(2) & 1.0 \\
  & & 2.0 & 24 & 2.60(3) & 0.7 \\
  & & 1.8 & 24 & 2.63(3) & 0.5 \\
  & & 1.6 & 24 & 2.62(3) & 3.6
  \\ \hline
  
$\Delta(\Omega,10,20,L)$ & (a) & 1.4  & 24  &  2.582(2)  & 1.8  \\
         &       &      & 32  &  2.582(4)  & 2.4  \\
         &       &  1.3  & 24  &  2.584(2)  & 3.6  \\
         &       &  1.2  & 24  &  2.582(4)  & 1.5  \\
  \hline

  $t(\Omega,40,L)$ & (a) & 2.2 & 32 & 2.536(2) & 14 \\
  & &
  & 40 & 2.552(2) & 0.3 \\
  & & & 48 & 2.549(12) & 0.2 \\
  & & 1.8 & 32 & 2.553(4) & 10 \\
  & & & 40 & 2.568(4) & 0.1 \\
  & & & 48 & 2.571(11) & 0.1 \\
  & & 1.6 & 32 & 2.566(2) & 5.2 \\
  & & & 40 & 2.569(3) & 6.0 \\
  & & & 48 & 2.576(6) & 10 \\
  \hline

  $C_2$  & (a) & $_{2.2,1.7}$ & 32  &  2.531(2)  & 13  \\
         &       &         & 40  &  2.551(3)  & 1.3  \\
         &       &         & 48  &  2.561(8)  & 0.7  \\
  \hline

  $C_{3}$ & (a) & $_{2.2,1.7,1.3}$ & 40  &  2.550(3)  & 5.7  \\
  &       &       & 48  &  2.566(8)  & 5.8  \\
          &        &              & 56  &  2.598(17)  & 11  \\
  \hline
  
\end{tabular}
\caption{Results for the dynamic exponent $z$ obtained by some fits of
  the data of $\Delta(\Omega,\Upsilon_1,\Upsilon_2,L)$ and
  $t(\Omega,\Upsilon,L)$ to the ansatz: (a) $a \, L^z$ and (b) $a
  \,L^z (1 + a_1 \,L^{-\omega})$ for the lattice ${\mathbb Z}_8$ gauge
  model, at fixed values of $\Omega$, using data for $\Upsilon=40$,
  $\Upsilon=20$, and $\Upsilon=10$. We also report combined fits $C_2$
  of the data of the data of $t(\Omega,\Upsilon,L)$ for $\Upsilon=40$,
  $\Upsilon=20$ (with $\Omega$ chosen so that they correspond to
  $\Theta\approx 0.02$) and combined fit $C_3$ for
  $\Upsilon=40,\,20,\,10$ (with $\Omega$ chosen so that they
  correspond to $\Theta\approx 0.02$). On the basis of these results,
  one may consider $z=2.590(35)$ as the optimal estimate the lattice
  ${\mathbb Z}_8$ gauge model, where the central value is taken from
  the most stable fits of the differences
  $\Delta(\Omega,\Upsilon_1,\Upsilon_2,L)$, which are apparently
  subject to smaller scaling corrections, and the error allows for the
  spread of the results of the fits.}
  \label{fitsz8}
\end{table}

We now present numerical MC results for the 3D lattice ${\mathbb Z}_6$
and ${\mathbb Z}_8$ gauge models up to lattice size $L=64$, focusing
on the out-of-equilibrium behavior of the subtracted energy density
$E_s(t,r,L)$, cf. Eq.~(\ref{diffet}), along the critical relaxational
flow driven by standard Metropolis upgradings, described in
Sec.~\ref{locrelmetro}.  We recall that one MC time unit corresponds
to a global sweep of a single Metropolis upgrading proposal for each
bond variable.

We analyze the data within the out-of-equilibrium FSS framework
outlined in Sec.~\ref{enesca}. We report results at fixed values of
$\Upsilon=r L^{y_r}$ with $r=K_c-K$ and $y_r^{-1}=\nu=0.6717$ (we
recall that $K<K_c$ is the coupling of the initial Gibbs ensemble of
configurations). In practice, along the equilibrium run at $K<K_c$ (we
also use the Metropolis algorithm to generate the starting equilibrium
configurations at $K<K_c$), corresponding to a given value of
$\Upsilon$, we start a critical relaxational trajectory every $n
\approx 0.2 \,L^z$ sweeps at equilibrium.  This distance between
trajectories provides a reasonable compromise to get almost
decorrelated starting configurations, as checked by preliminary tests
with the approximate estimate $z\approx 2.6$.  The data for the
time-dependent energy density are obtained by averaging over a large
number of trajectories.  We typically collected $O(10^5)$ critical
relaxational trajectories for the lattice sizes up to $L=48$ and
$O(10^4)$ trajectories for the largest lattices $L=56,\,64$.  The
error on the average of the energy density over the trajectories is
estimated by a standard Whitmer blocking
procedure~\cite{Whitmer-84,GMTW-86,FP-89}, to properly take into
account the residual autocorrelations among the sequential
trajectories along the equilibrium run.

To estimate the dynamic exponent $z$, we perform fits of the data of
$t(\Omega,\Upsilon,L)$, cf. Eq.~(\ref{generalansatzt}), and
$\Delta(\Omega,\Upsilon_1,\Upsilon_2,L)$, cf. Eq.~(\ref{difft}), at
fixed values of $\Upsilon$ and $\Omega$. Note that these sets of data
are statistically independent, so that their fits avoid the problem of
the highly correlated data along the critical relaxational flow.

\begin{figure}[tbp]
  \includegraphics[width=0.9\columnwidth, clip]{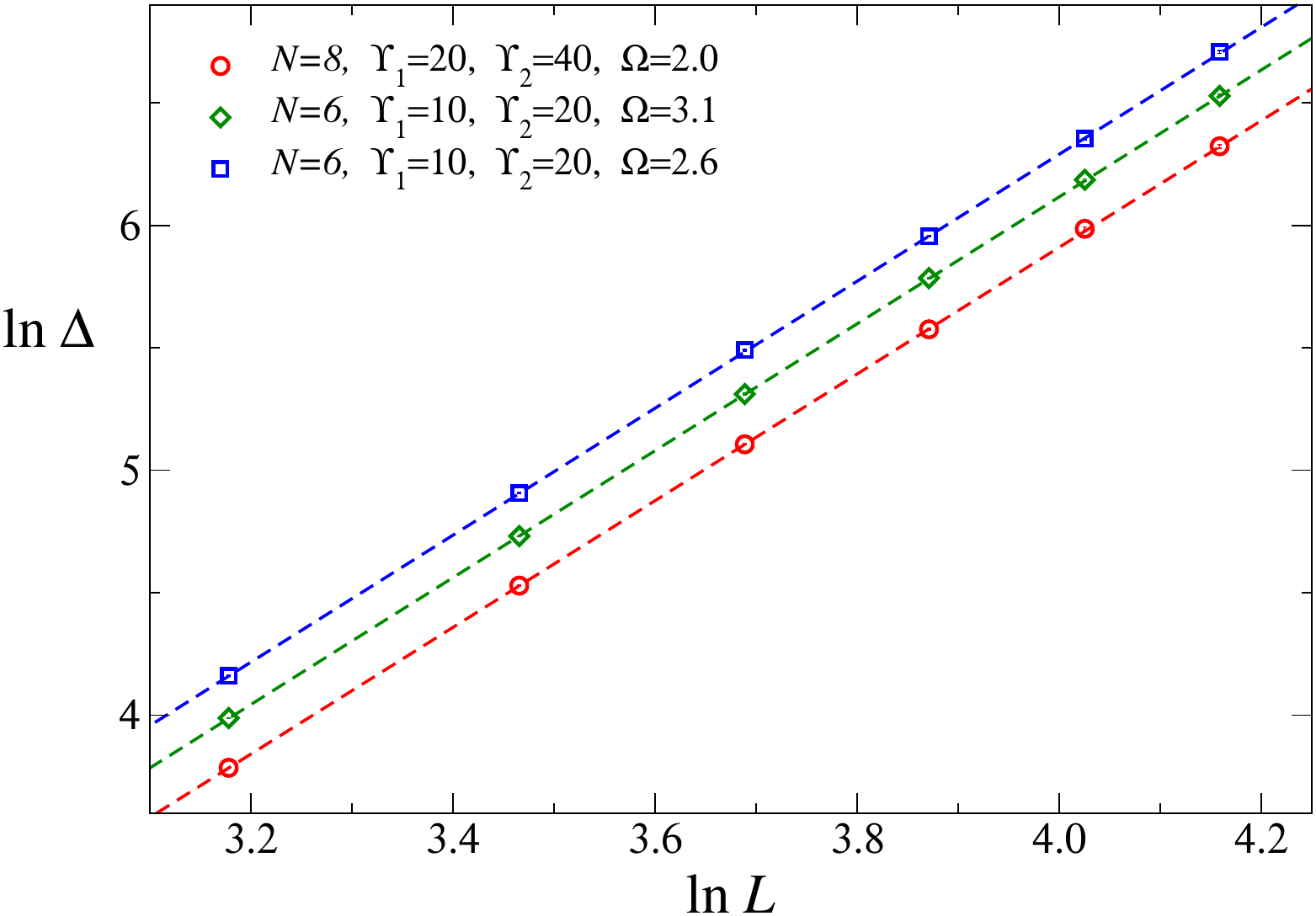}
  \caption{We show a log-log plot of some data of the difference
    $\Delta(\Omega,\Upsilon_1,\Upsilon_2,L)$, defined in
    Eq.~(\ref{difft}), versus the lattice size up to $L=64$, for $N=6$
    and $N=8$ and some values of $\Omega$, $\Upsilon_1$ and
    $\Upsilon_2$. The dashed lines show their best fits to $a\, L^z$
    with $z=2.59$, to highlight the accuracy of the estimate of $z$.
    Analogous results are obtained for other values of $\Omega$ (for
    example within the range corresponding to $0.005\lesssim \Theta
    \lesssim 0.04$) and $\Upsilon$.
  }
    \label{dtomega}
\end{figure}

In Tables \ref{fitsz6} and \ref{fitsz8}, we report the results of some
fits at some values of $\Omega$, corresponding to the optimal range
$0.01\lesssim \Theta \lesssim 0.04$, most of them to the simplest
power-law ansatz $f(L) = a\, L^z$ predicted by the out-of-equilibrium
FSS theory. The fits of the differences
$\Delta(\Omega,\Upsilon_1,\Upsilon_2,L)$ provide the most accurate
results, because they turn out to be much less affected by scaling
corrections.  This can be easily checked by verifying the stability of
the results of the fits to the simplest power-law ansatz $a \,L^z$,
when systematically increasing the minimum value $L_{\rm min}$ of the
data allowed in the fit.  Note that for the differences
$\Delta(\Omega,\Upsilon_1,\Upsilon_2,L)$ we generally obtain stable
results starting from $L_{\rm min}=24$, with acceptable $\chi^2/{\rm
  d.o.f}\lesssim 1$. In Tables~\ref{fitsz6} and \ref{fitsz8} we also
report some fits to the ansatz $a \,L^z (1 + a_1 \,L^{-\omega})$
including the expected leading scaling corrections. They give slightly
larger results, but fully compatible within the errors (they are
useful to estimate the error of the estimate of $z$).  On the other
hand, the fits of the data of $t(\Omega,\Upsilon,L)$ show clearly
evidence of sizeable scaling corrections, requiring larger values of
$L_{\rm min}$ to achieve an acceptable value of $\chi^2/{\rm d.o.f}$.
We also mention that fits of the data of $t(\Omega,\Upsilon,L)$,
including the $O(L^{-\omega})$ scaling corrections, do not provide
sufficiently stable results (larger lattice sizes are required to
stabilize them).

The results obtained by the above analyses, see in particular the
results reported in Tables~\ref{fitsz6} and ~\ref{fitsz8}, lead us to
the estimates $z=2.595(35)$ for the ${\mathbb Z}_6$ gauge model and
$z=2.590(35)$ for the ${\mathbb Z}_8$ gauge model.  Their errors take
into account the range of results obtained by the different fits
considered (the statistical errors obtained from the fits are
generally much smaller), and also the uncertainty on the available
estimates of $K_c$ and $E_{c,\infty}$, whose contributions to the
global error are much smaller.  They turn out to be in full agreement,
confirming the universality of the relaxational dynamic exponent $z$
within systems belonging to the IXY universality class.

\begin{figure}[tbp]
  \includegraphics[width=0.9\columnwidth, clip]{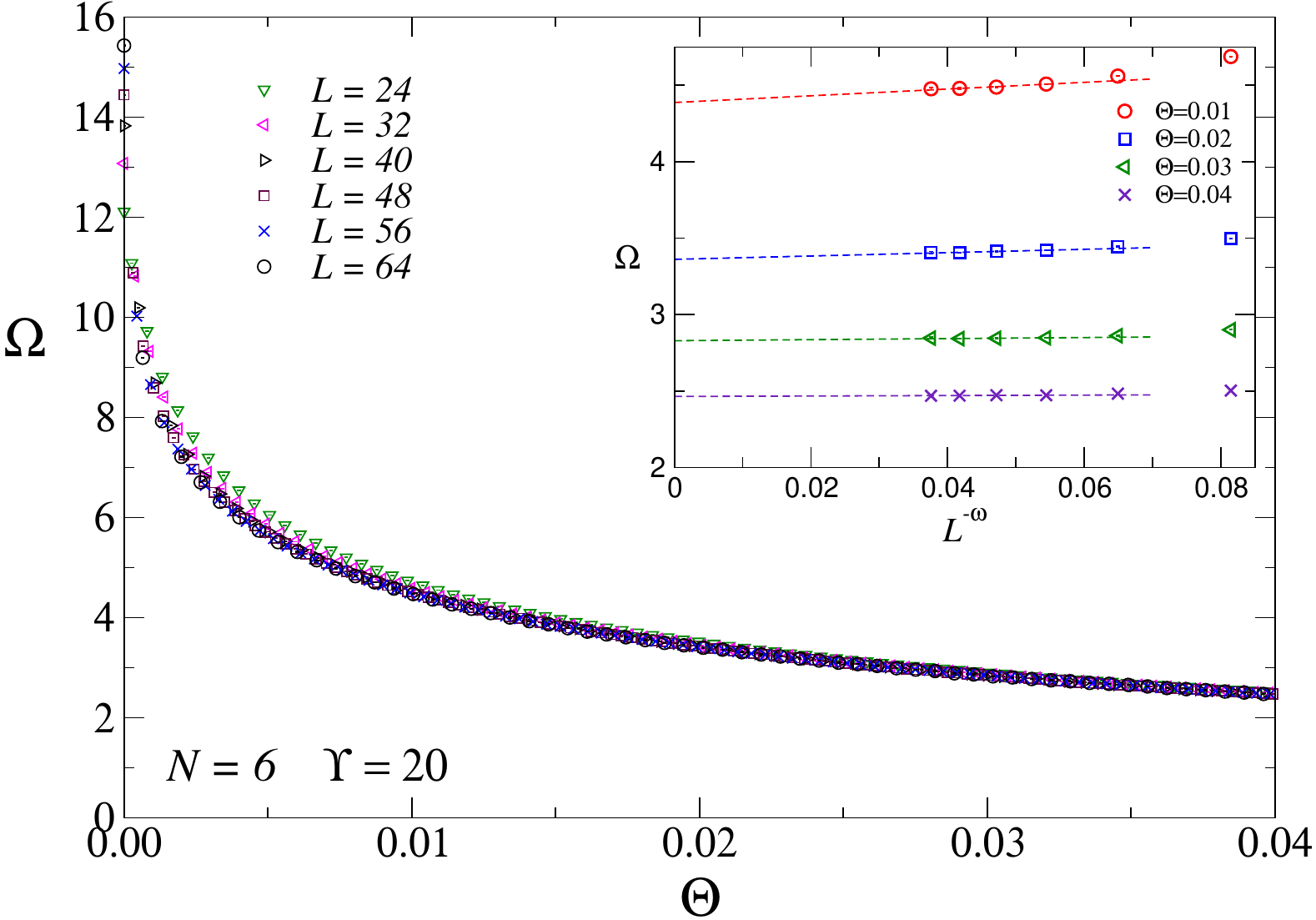}
    \includegraphics[width=0.9\columnwidth, clip]{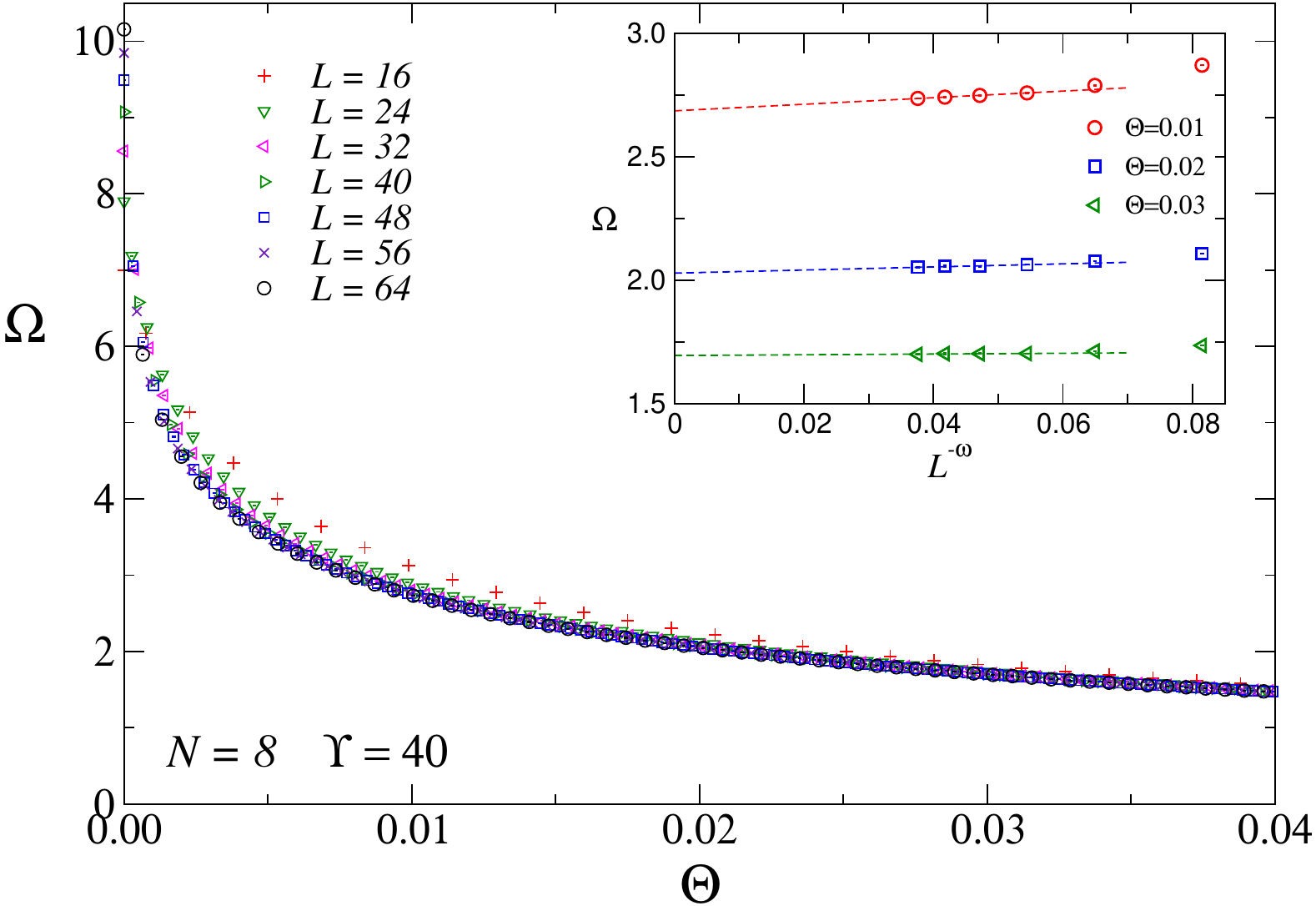}
  \caption{Out-of-equilibrium FSS along the relaxational flow of the
    subtracted energy density, and in particular of the scaling
    function $\Omega(t,r,L)=L^{3-y_r} E_s(t,r,L)$, versus
    $\Theta=t/L^z$, using our optimal estimate $z=2.59$.  We plot data
    for $N=6$ and $\Upsilon=20$ (top) and for $N=8$ and $\Upsilon=40$
    (bottom).  The statistical errors of the data are very small and
    hardly visible in these plots.  The insets show the convergence to
    the asymptotic scaling behavior for some values of $\Theta$,
    plotting the data of $\Omega$ versus $L^{-\omega}$ which is the
    expected rate of convergence (the dashed lines show linear fits of
    the data at fixed $\Theta$ to $a + b L^{-\omega}$ for $L\ge 40$).
    Analogous results are obtained for the other values of
    $\Upsilon$.}
    \label{OmegavsTheta}
\end{figure}

Therefore, assuming universality, We consider
\begin{equation}
  z = 2.59(3)
  \label{zestimate}
\end{equation}
as our final estimate for the dynamic exponent $z$ associated with the
relaxational dynamics of the IXY universality class.  The quality of
the results can be appreciated by looking at the plots reported in
Fig.~\ref{dtomega}, which shows some sets of data of the time
difference $\Delta(\Omega,\Upsilon_1,\Upsilon_2,L)$ for $N=6$ and
$N=8$, with their power-law fits. To further demonstrate the accuracy
of the estimate of $z$, in Figs.~\ref{OmegavsTheta} and we show plots
of $\Omega(t,r,L)$ versus $\Theta=t/L^z$ using the estimate
(\ref{zestimate}): the asymptotic collapse of the curves for any
$\Theta>0$ is clearly observed, nicely confirming the scaling ansatz
(\ref{erscal}).

\section{Equilibrium critical dynamics}
\label{equicritdyn}

In this section we report a numerical study of the critical
relaxational dynamics at equilibrium for the lattice ${\mathbb Z}_6$
gauge model.  For this purpose we study the autocorrelation functions
of gauge-invariant observables along equilibrium MC simulations at the
critical point. We still consider the local relaxational dynamics
realized by standard Metropolis upgradings of the link variables, see
Sec.~\ref{locrelmetro}.  The autocorrelation times, and in particular
their asymptotic large-$L$ power-law divergence, provide information
on the critical slowing down of the time scale $\tau$ of the critical
modes, which is expected to increase as $\tau\sim L^z$ with increasing
$L$.  Thus they allow us to estimate the dynamic critical exponent $z$
controlling the critical slowing down of the system under local
relaxational dynamics.

The autocorrelation function $C_o(t)$ along the equilibrium
relaxational dynamics of a given quantity $O$ is defined as
\begin{equation}
C_o(t) = \langle \left( O(t+t_0) - \langle O \rangle \right)
\left( O(t_0) - \langle O \rangle \right) \rangle,
\end{equation}
where $t,t_0$ are discrete times along the equilibrium MC evolutions,
and the averages are taken at equilibrium.  We focus on the integrated
autocorrelation time $\tau_o$ associated with $O$ is given by
\begin{equation}
  \tau_o = {1\over 2} \sum_{t=-\infty}^{+\infty} {C_o(t)\over C_o(0)}=
      {1\over 2} + \sum_{t=1}^{+\infty} {C_o(t)\over C_o(0)}.
      \label{integrauto}
\end{equation}
The integrated autocorrelation times can be estimated more easily than
the time $\tau_{\rm exp}$ controlling the exponential large-time
decays of the autocorrelation functions, i.e.,
\begin{equation}
C_o(t)\sim e^{-t/\tau_{\rm exp}}.
\label{expbeh}
\end{equation}
In particular, an accurate estimate of the exponential-decay time
$\tau_{\rm exp}$ requires very long equilibrium MC runs, many
times the typical autocorrelation times, generally requiring a much
larger computational effort than that necessary to reliably estimate
the integrated autocorrelation time.

Estimates of $\tau_o$ can be obtained by the binning
method~\cite{Wolff-04,DMV-04}, using the estimator
\begin{equation}
\tau_o = {A_o^2\over 2 A_{o,n}^2},
\label{eqbl}
\end{equation}
where $A_{o,n}$ is the naive error calculated without taking into
account the autocorrelations, and $A_o$ is the correct error found
after binning, i.e. when the error estimate becomes stable with
respect to increasing of the block size $b$.  The statistical error
$\Delta\tau_o$ is just obtained by the ratio $\Delta \tau_o/\tau_o =
\sqrt{2/n_{b}}$ where $n_{b}$ is the number of blocks corresponding to
the estimate of $A_o$.  As discussed in Ref.~\cite{Wolff-04} this
procedure leads to a systematic error of $O(\tau_o/b)$.  In our cases
the ratio $\tau_o/b$ will always be smaller than the statistical
error, so we will neglect it.

In our study of the equilibrium dynamics we focus on the integrated
autocorrelation time $\tau_p$ of the Polyakov loops, defined as the
product of all link variables along one direction, such as
\begin{equation}
P_{\bm x} = \prod_{k=0}^{L-1}\lambda_{{\bm x}+k\hat{\mu},\mu}\ .
\label{poly}
\end{equation}
$P_{\bm x}$ is gauge invariant due to the periodic boundary
conditions. Actually we consider its spatially (zero-momentum)
integral, which provides more stable results. This nonlocal observable
is expected to have a substantial overlap with the critical modes
associated with the topological deconfinement transition. Its
autocorrelation time turns out to be quite large, being $\tau_p\gtrsim
10^5$ for $L\ge 24$ [the typical number of global sweeps collected in
  our runs is $O(10^9)$].

\begin{figure}[tb]
  \includegraphics[width=0.9\columnwidth, clip]{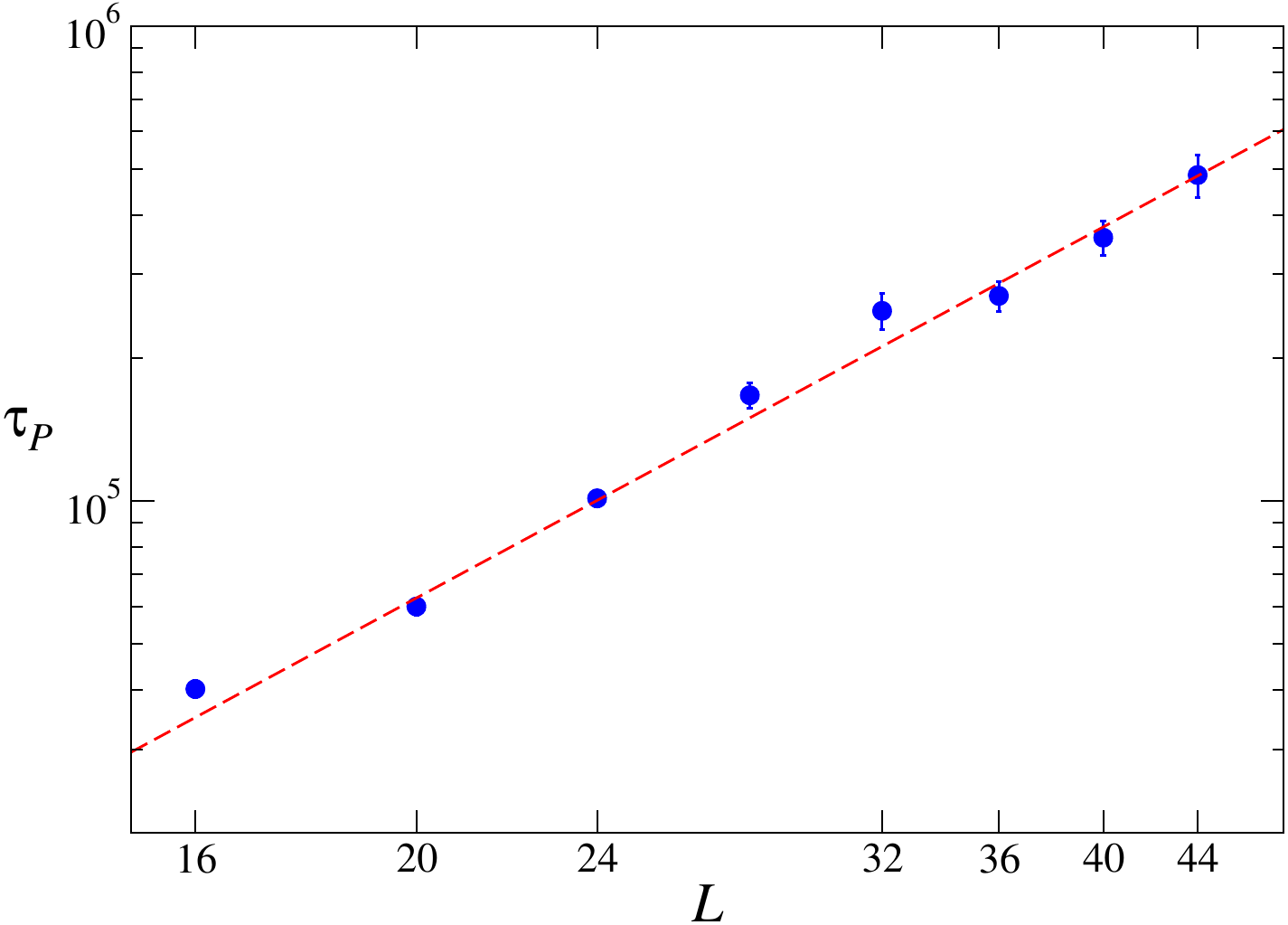}
  \caption{Integrated autocorrelation time of the Polyakov loop in the
    $\mathbb{Z}_6$ model at criticality.  The dashed line is the
    result of a fit of the data for $L\ge 20$ to the power law
    $\tau_p=a\,L^z$ with fixed $z=2.59$.}
    \label{tauintpoly}
\end{figure}

The data of $\tau_p$ are plotted in Fig.~\ref{tauintpoly} versus the
size $L$, up to $L=44$.  Their behavior appears nicely consistent with
the expected power law $a\, L^z$ using the estimate $z=2.59$ obtained
from the analyses of the out-of-equilibrium relaxational flows.  An
unbiased analysis of the data of $\tau_p$, by fitting also the dynamic
exponent of the ansatz $a\,L^z$, gives the independent estimate
$z=2.52(12)$, which is fully consistent with our best estimate
$z=2.59(3)$.

We recall that all observables coupled to the slowest critical modes
are expected to show an analogous large-time exponential decay with
$\tau_{\rm exp}\sim L^z$.  As already remarked in
Ref.~\cite{Sokal-97}, the dynamic critical exponent governing the
divergence of the integrated autocorrelation time $\tau_o\sim
L^{z_{i}}$ (for a given observable $O$) may be smaller than that
characterizing the asymptotic behavior of the autocorrelation time
$\tau_{\rm exp}$ associated with the evolution of the slowest modes of
the Markov chain, i.e. $z_i\le z$. This may in particular occur when
the observable considered does not have a substantial coupling with
the slowest long-distance equilibrium critical modes. Therefore, when
focusing on the integrated autocorrelation times, one should take the
maximum value $z_i$ among the various observables considered as
estimate of the global dynamic exponent $z$.  In this respect, it is
worth mentioning that a smaller value $z_i\approx 2.3$ was apparently
obtained by using the integrated autocorrelation time of the energy
density. On the other hand, by estimating the autocorrelation time
$\tau_{\rm exp}$ from the MC history of the energy density (using the
same technique adopted in \cite{BPV-25-z2equi}), we obtained a larger
result for $z$, fully compatible with our best estimate $z=2.59(3)$,
but with a significantly larger error.

\section{Conclusions}
\label{conclu}

We report a numerical study of the critical relaxational dynamics of
3D lattice ${\mathbb Z}_N$ gauge models defined by the Hamiltonian
(\ref{zqgau}), for $N=6$ and $N=8$, which both undergo topological
continuous transitions belonging to the 3D IXY universality class.
The static critical behavior of the lattice ${\mathbb Z}_N$ gauge
model is related by duality to that of the ${\mathbb Z}_N$ clock model
with a global ${\mathbb Z}_N$ symmetry~\cite{Savit-80,BCCGPS-14},
implying that they share an analogous critical behavior in the thermal
sector, and, in particular, the same length-scale critical exponent
$\nu$.  Nevertheless, we show that the critical relaxational dynamics
of lattice ${\mathbb Z}_N$ gauge models turns out to be significantly
slower than that of the corresponding spin systems.

The purely relaxational dynamics is realized by a local reversible
Metropolis dynamics.  Our study allows us to determine the
relaxational dynamic exponent $z$ controlling the universal power law
of the critical slowing down at the topological transitions belonging
to the 3D IXY universality class.  For this purpose, we focus on the
out-of-equilibrium dynamics along the relaxational flows arising from
instantaneous quenches to the critical point, and analyze them within
an out-of-equilibrium FSS framework.  The analyses of the
out-of-equilibrium relaxational flows for the lattice ${\mathbb Z}_6$
and ${\mathbb Z}_8$ models provide consistent results, leading to our
final estimate $z=2.59(3)$. This result is also supported by some
analyses of the equilibrium critical dynamics of the lattice ${\mathbb
  Z}_6$ model at the critical point, obtaining consistent, although
less accurate, results.

We conjecture that the dynamic exponent
$z=2.59(3)$ characterizes the critical slowing down of the relaxational
dynamics at all topological transitions belonging to the IXY
universality class, such as the lattice IXY gauge model (\ref{IXYham})
and the lattice Abelian Higgs model which provides an effective theory
for the continuous transitions in superconductors, see, e.g.,
Refs.~\cite{BPV-24-rev,DH-81,NRR-03,BPV-24-decQ2,BPV-24-ncAH,BPV-21-ncAH,
  HLM-74,HT-96,FH-96,BFLLW-96,OT-98,HS-00,KNS-02,MHS-02,Herbut-book,BPV-23b}.

These results imply that the power law of the relaxational critical
slowing down of models belonging to the 3D IXY universality class
differs from that of the standard XY universality class, although
their partition functions are exactly related by duality.  Indeed, the
dynamic exponent $z$ of the IXY universality class turns out to be
larger than the dynamic exponent $z\approx 2.02$ associated with the
purely relaxational dynamics of the standard 3D XY universality
class~\cite{AEHIKKZ-22,FM-06,AV-84,HH-77,HHM-72,BPV-25-z2equi}.  In this
respect, it is important to note that the duality
mapping~\cite{DH-81,NRR-03} between the IXY gauge model and the XY
model with Villain action, which guarantees the equivalence of the
equilibrium critical behavior in the thermal sector, is
nonlocal. Therefore, a local dynamics in the XY model would correspond
to a nonlocal dynamics in the IXY model, and vice versa, giving rise
to different dynamic universality classes for the IXY gauge model and
the standard XY model.  An analogous scenario characterizes the local
relaxational dynamics of the 3D ${\mathbb Z}_2$ gauge model, i.e. the
model (\ref{zqgau}) with $N=2$, whose equilibrium behavior exactly
maps into that of the standard 3D Ising model by duality, with dynamic
exponents $z=2.610(15)$~\cite{BPV-25,BPV-25-z2equi} and
$z=2.0245(15)$~\cite{Hasenbusch-20}, respectively.  We also note that
estimates of the dynamic relaxational exponents of the ${\mathbb Z}_2$
gauge and IXY universality classes are numerically close.

The above results for the lattice ${\mathbb Z}_N$ gauge theories and
the inverted XY universality class suggest a general trend of the
critical dynamics of lattice gauge systems: The critical slowing down
arising from local relaxational dynamics in lattice gauge theories is
generally stronger than that of spin systems without gauge symmetries,
in particular with respect to those whose static behaviors are related
by duality.

\begin{acknowledgments}
  
H.P. thanks the University of Pisa for the kind hospitality. H.P. would
like to acknowledge support from the Cyprus Research and Innovation
Foundation (RIF).

\end{acknowledgments}





\end{document}